\documentclass[12pt]{article}
\usepackage[utf8x]{inputenc}
\usepackage[russian]{babel}
\usepackage{a4wide,amsmath,graphicx}
\numberwithin{equation}{section}

\begin{document}
\begin{flushright}
\begin{minipage}{0.7\textwidth}
\begin{center}
\includegraphics[width=4cm]{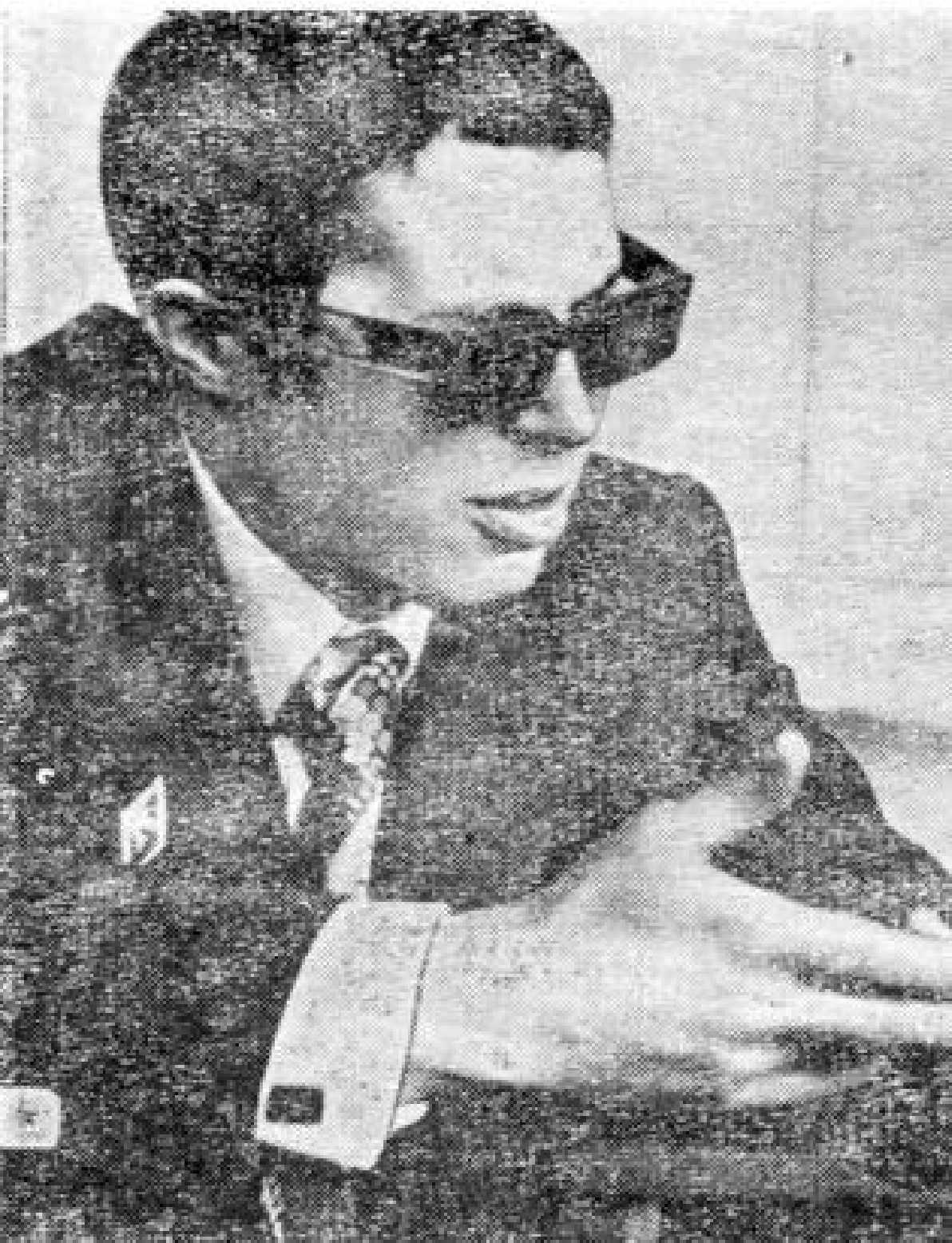}
\end{center}
Светлой памяти Владимира Александровича Жегалина (1947--2002) ---
учителя физики, который умел заинтересовать своих учеников,
и тем самым определил мою дальнейшую жизнь.
\end{minipage}
\end{flushright}
\vspace{5mm}
\begin{center}
{\LARGE Специальная теория относительности}\\[5mm]
Андрей Грозин\\[5mm]
\end{center}
\begin{abstract}
В этой статье кратко и (я надеюсь) доступно излагаются
основные положения специальной теории относительности.
Я старался показать, что теория относительности
не только не противоречит здравому смыслу,
но, наоборот, логически следует из него.
Подробно обсуждается геометрия пространства--времени Минковского.
Геометрический подход (с минимумом формул) позволяет сделать
выводы теории относительности наглядными и интуитивно очевидными.
Я использовал радиолокационный подход Бонди,
который позволяет прийти к выводам теории кратчайшим путём.

Использовать скорость света $c$ в теории относительности ---
это всё равно что изучать евклидову геометрию,
измеряя координату $x$ в сантиметрах, а $y$ в дюймах.
Во всех формулах будет назойливо присутствовать
``фундаментальная константа'' $c = 2.54\;$см/дюйм;
естественность и очевидность результатов полностью исчезнут.
\end{abstract}

\section{Пространство--время}
\label{S:st}

Пространство--время --- это множество всех событий.
Событие характеризуется тем, что произошло где-то и когда-то.
Оно однозначно определяется четырьмя числами ---
моментом времени и тремя пространственными координатами.
То есть пространство--время четырёхмерно.
Все события, произошедшие с частицей, образуют её мировую линию.
В каждый момент времени частица где-то находилась,
а это уже событие.
Мировая линия представляет собой одномерное подмножество
пространства--времени.

\begin{figure}[ht]
\begin{center}
\begin{picture}(52,53)
\put(25,28){\makebox(0,0){\includegraphics{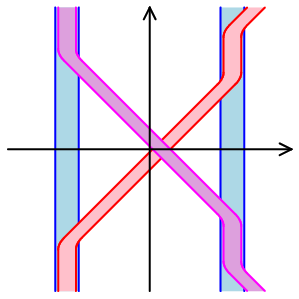}}}
\put(23,52){\makebox(0,0){$t$}}
\put(49,26){\makebox(0,0){$x$}}
\put(11,0){\makebox(0,0)[b]{Станция 1}}
\put(39,0){\makebox(0,0)[b]{Станция 2}}
\end{picture}
\end{center}
\caption{Движение поезда описывается двумерной областью между мировыми линиями его головы и хвоста.}
\label{F:Metro}
\end{figure}

Рисовать четырёхмерные картинки довольно неудобно,
поэтому мы будем в основном рассматривать движение частиц вдоль прямой.
Тогда пространство--время двумерно, и его легче себе представить.
Например, на рис.~\ref{F:Metro} показано движение поездов метро.

Каждый наблюдатель имеет при себе часы,
и может измерять время событий на своей мировой линии.
Мы примем постулат, что существуют правильно идущие часы.
Показания двух правильных часов одного наблюдателя
могут отличаться лишь выбором начала отсчёта
(естественно, предполагается, что тиканье разных часов
приведено к единой единице измерения времени).
Вопрос о том, как сделать правильные часы,
относится не к физике пространства--времени,
а к физике населяющей его материи.
Он кратко обсуждается в приложении~\ref{S:clock}
(необязательном для понимания основного текста).
Наблюдатель не может непосредственно измерить время события
вне своей мировой линии.
Этот процесс неизбежно включает в себя передачу сигналов,
и будет подробно рассмотрен в \S~\ref{S:L}.

\begin{figure}[h]
\begin{center}
\begin{picture}(23,49)
\put(13,26){\makebox(0,0){\includegraphics{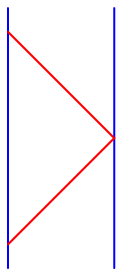}}}
\put(0,8){\makebox(0,0)[l]{$t_1$}}
\put(0,44){\makebox(0,0)[l]{$t_2$}}
\put(4,0){\makebox(0,0)[b]{1}}
\put(22,0){\makebox(0,0)[b]{2}}
\end{picture}
\end{center}
\caption{Обмен сигналами.}
\label{F:sig1}
\end{figure}

Допустим, наблюдатель 1 посылает сигнал наблюдателю 2 в момент $t_1$,
а тот немедленно посылает сигнал обратно (рис.~\ref{F:sig1}).
Логически, можно представить себе три возможности:
\begin{itemize}
\item Возможна ситуация, когда $t_2<t_1$, то есть ответный сигнал
приходит раньше отправления первого сигнала.
К счастью, эта возможность не соответствует нашему реальному миру;
иначе про причинность можно было бы забыть.
\item Разность $t_2-t_1$ положительна, но может быть сделана сколь угодно малой
путём выбора подходящих сигналов и технических усовершенствований.
В этом случае можно ввести абсолютное время, одинаковое для всех наблюдателей.
Именно так все и думали до создания теории относительности.
\item Разность $t_2-t_1$ не может быть сделана меньше некоторого положительного значения.
То есть существует наибыстрейший сигнал.
Как показывают эксперименты, именно так устроен реальный мир.
\end{itemize}

Таким наибыстрейшим сигналом является свет (а также гравитационные волны).
Это связано с тем, что фотоны (и гравитоны) --- безмассовые частицы (\S~\ref{S:p}).
Однако, конкретная природа наибыстрейших сигналов
не является важной для теории относительности.
Если когда-нибудь у фотона обнаружат очень маленькую ненулевую массу,
то это будет серьёзным потрясением для теории электромагнетизма,
но никак не затронет основ теории относительности.
Важен только сам факт, что никакой сигнал не может быть быстрее чего-то.
Мы будем для определённости называть наибыстрейшие сигналы световыми.

\begin{figure}[h]
\begin{center}
\begin{picture}(50,36)
\put(25,18){\makebox(0,0){\includegraphics{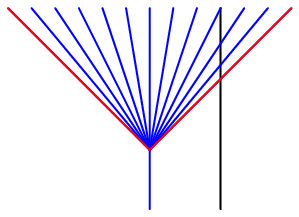}}}
\put(23,9){\makebox(0,0){$O$}}
\put(39,21){\makebox(0,0){$A$}}
\end{picture}
\end{center}
\caption{Взрыв бомбы.}
\label{F:bomb}
\end{figure}

Пусть событие $O$ будет взрывом бомбы.
Тогда наблюдатель сначала увидит вспышку (событие $A$),
а потом уже на него посыплются осколки (рис.~\ref{F:bomb}).
Мировые линии световых сигналов, испущенных в точке $O$,
образуют световой полуконус будущего этого события.
Событие $O$ может влиять на события, лежащие внутри (и на границе)
этого полуконуса, при помощи света и других сигналов
(например, осколков).
Эта область пространства--времени называется \emph{будущим}
события $O$.

\begin{figure}[ht]
\begin{center}
\begin{picture}(42,42)
\put(21,21){\makebox(0,0){\includegraphics{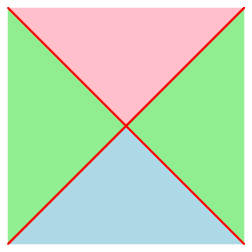}}}
\put(21,37){\makebox(0,0){будущее}}
\put(21,5){\makebox(0,0){прошлое}}
\put(41,21){\makebox(0,0)[r]{удалённое}}
\put(18,21){\makebox(0,0){$O$}}
\end{picture}
\end{center}
\caption{Будущее, прошлое и удалённое события $O$.}
\label{F:cone1}
\end{figure}

Аналогично, мировые линии световых сигналов, приходящих в точку $O$,
образуют световой полуконус прошлого этого события.
События внутри (и на границе) этого полуконуса могут влиять на событие $O$.
Эта область называется \emph{прошлым} события $O$.
Область пространства--времени вне светового конуса
называется \emph{удалённым} события $O$ (рис.~\ref{F:cone1}).
События из этой области не могут влиять на $O$,
и $O$ не может влиять на них.

\begin{figure}[ht]
\begin{center}
\begin{picture}(40,48)
\put(20,20){\makebox(0,0){\includegraphics[width=40mm]{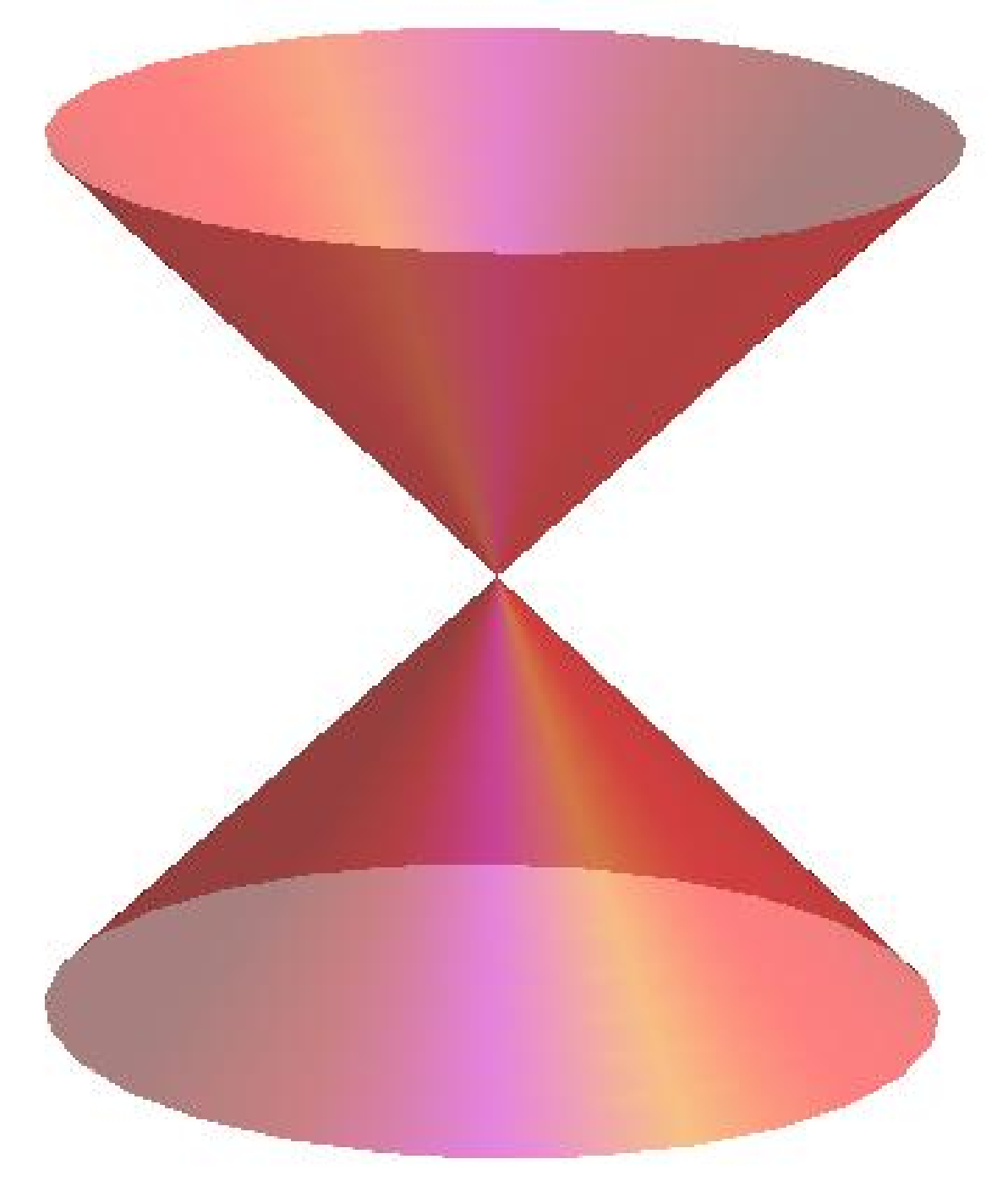}}}
\end{picture}
\end{center}
\caption{Световой конус.}
\label{F:cone}
\end{figure}

Световой конус в трёхмерном пространстве--времени (две пространственных координаты)
показан на рис.~\ref{F:cone}.
Будущее находится внутри светового полуконуса будущего;
прошлое --- внутри светового полуконуса прошлого;
удалённое --- вне светового конуса.
В четырёхмерном пространстве--времени световой конус ---
это трёхмерная поверхность, нарисовать которую труднее.

\section{Преобразования Лоренца}
\label{S:L}

Частица, на которую не действуют никакие силы, движется по инерции.
Принцип относительности Галилея гласит, что
\emph{все инерциальные наблюдатели равноправны}.
Если один инерциальный наблюдатель поставил какой-то эксперимент
и получил некоторый результат,
а другой инерциальный наблюдатель поставил такой же эксперимент,
то он получит такой же результат.

\begin{figure}[ht]
\begin{center}
\begin{picture}(25,51)
\put(12.5,24){\makebox(0,0){\includegraphics{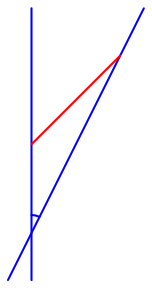}}}
\put(5,51){\makebox(0,0)[t]{1}}
\put(24,51){\makebox(0,0)[t]{2}}
\put(3,24){\makebox(0,0){$t_1$}}
\put(22,37){\makebox(0,0){$t_2$}}
\put(6.5,14){\makebox(0,0){$\varphi$}}
\put(2.5,8){\makebox(0,0){$O$}}
\end{picture}
\end{center}
\caption{Два инерциальных наблюдателя.}
\label{F:sig2}
\end{figure}

Допустим, два инерциальных наблюдателя пролетели мимо друг друга
(событие $O$, рис.~\ref{F:sig2}).
Их часы установлены в 0 в точке $O$.
Наблюдатель 1 послал световой сигнал в момент времени $t_1$ (по своим часам);
наблюдатель 2 его принял в момент $t_2$ (по своим часам).
Тогда
\begin{equation}
t_2 = t_1\,e^\varphi\,,
\label{L:k}
\end{equation}
где величина $\varphi$ называется углом между мировыми линиями
наблюдателей 1 и 2.

\begin{figure}[ht]
\begin{center}
\begin{picture}(29.6,51)
\put(14.8,24){\makebox(0,0){\includegraphics{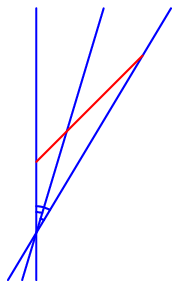}}}
\put(5.8,51){\makebox(0,0)[t]{1}}
\put(17.2,51){\makebox(0,0)[t]{2}}
\put(28.6,51){\makebox(0,0)[t]{3}}
\put(3.8,21){\makebox(0,0){$t_1$}}
\put(9,28){\makebox(0,0){$t_2$}}
\put(26,37.5){\makebox(0,0){$t_3$}}
\put(3,12){\makebox(0,0){$\varphi_{12}$}}
\put(10.5,10.5){\makebox(0,0){$\varphi_{23}$}}
\put(8,16){\makebox(0,0){$\varphi_{13}$}}
\end{picture}
\end{center}
\caption{Три инерциальных наблюдателя в одной плоскости.}
\label{F:sig3}
\end{figure}

Почему эту величину естественно назвать углом?
Рассмотрим трёх инерциальных наблюдателей,
движущихся в одной плоскости.
Пусть их мировые линии пересекаются в одной точке
($O$, рис.~\ref{F:sig3}).
Тогда
\begin{equation*}
t_2 = t_1\,e^{\varphi_{12}}\,,\qquad
t_3 = t_1\,e^{\varphi_{13}} = t_2\,e^{\varphi_{23}}\,,
\end{equation*}
откуда следует
\begin{equation}
\varphi_{13} = \varphi_{12} + \varphi_{23}\,,
\label{L:phi}
\end{equation}
как в обычной евклидовой геометрии.
Подчеркнём ещё раз, что простой закон сложения углов~(\ref{L:phi})
верен только для мировых линий в одной плоскости
(как и в евклидовой геометрии).

\begin{figure}[ht]
\begin{center}
\begin{picture}(32.5,66)
\put(16.25,31.5){\makebox(0,0){\includegraphics{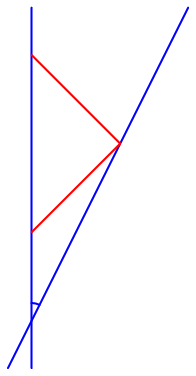}}}
\put(5,66){\makebox(0,0)[t]{1}}
\put(31.5,66){\makebox(0,0)[t]{2}}
\put(3,24){\makebox(0,0){$t_1$}}
\put(3,54){\makebox(0,0){$t_2$}}
\put(21.5,39){\makebox(0,0){$t$}}
\put(17,39){\makebox(0,0){$A$}}
\put(2.5,8){\makebox(0,0){$O$}}
\put(6.5,14){\makebox(0,0){$\varphi$}}
\end{picture}
\end{center}
\caption{Обмен сигналами между двумя инерциальными наблюдателями.}
\label{F:l}
\end{figure}

Пусть теперь наблюдатель 1 посылает световой сигнал в момент $t_1$ (по своим часам);
наблюдатель 2 получает его в момент $t$ (событие $A$),
и сразу посылает ответный сигнал;
наблюдатель 1 получает его в момент $t_2$ (рис.~\ref{F:l}).
Как мы знаем, $t=t_1\,e^{\varphi}$;
кроме того, $t_2=t\,e^{\varphi}$, ведь наблюдатели 1 и 2 равноправны.
Значит,
\begin{equation}
t_1 = t\,e^{-\varphi}\,,\qquad
t_2 = t\,e^{\varphi}\,.
\label{L:t12}
\end{equation}

Событие $A$ произошло вне мировой линии наблюдателя 1,
поэтому он не может непосредственно измерить время этого события по своим часам.
Наиболее естественно определить время события $A$ с точки зрения наблюдателя 1
как середину интервала $[t_1,t_2]$,
так как световой сигнал распространяется в обе стороны одинаково быстро.
По той же причине, наиболее естественно определить координату события $A$
с точки зрения наблюдателя 1 как половину этого промежутка времени.
Скорость наибыстрейшего сигнала мы принимаем за 1,
что фиксирует естественную единицу измерения расстояния;
сигнал пропутешествовал от наблюдателя 1 до события $A$ и обратно.
Таким образом, по определению, временная и пространственная координаты
события $A$ с точки зрения наблюдателя 1 есть
\begin{equation}
x^0 = \frac{t_1+t_2}{2}\,,\qquad
x^1 = \frac{t_2-t_1}{2}\,.
\label{L:x01}
\end{equation}
Подставляя~(\ref{L:t12}), получим
\begin{equation}
x^0 = t \cosh\varphi\,,\qquad
x^1 = t \sinh\varphi\,.
\label{L:csh}
\end{equation}

Мы пришли к очень важному результату: величина
\begin{equation}
x^2 \equiv (x^0)^2 - (x^1)^2 = t^2
\label{L:x2}
\end{equation}
не зависит от того, как движется наблюдатель 1
(т.\,е.\ от $\varphi$).
Компоненты $x^0$, $x^1$ вектора $x$ (из точки $O$ в точку $A$)
различны для разных наблюдателей;
инвариантную величину $x^2$ естественно назвать квадратом длины вектора $x$,
т.\,е.\ квадратом расстояния от $O$ до $A$.
Это расстояние есть $t$, то есть интервал времени между событиями $O$ и $A$
по часам наблюдателя 2 (оба события лежат на его мировой линии).
Формула~(\ref{L:x2}) для квадрата длины вектора отличается
от привычной евклидовой тем, что вместо знака $+$ между двумя членами
стоит знак $-$.
Поэтому геометрия пространства--времени называется псевдоевклидовой
(или геометрией Минковского).

Из формулы~(\ref{L:csh}) видно, что привычная ньютоновская скорость
наблюдателя 2 по отношению к наблюдателю 1 есть
\begin{equation}
u = \frac{x^1}{x^0} = \tanh \varphi
\label{L:u}
\end{equation}
(она всегда $<1$, и стремится к 1 при $\varphi\to\infty$).
Эта величина, однако, неудобна;
удобнее использовать угол $\varphi$ между мировыми линиями,
обладающий естественным свойством аддитивности~(\ref{L:phi}).

\begin{figure}[h]
\begin{center}
\begin{picture}(22.4,69)
\put(11.2,33){\makebox(0,0){\includegraphics{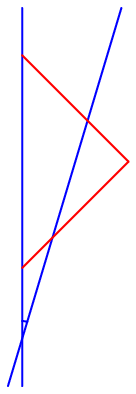}}}
\put(3.4,69){\makebox(0,0)[t]{1}}
\put(20.2,69){\makebox(0,0)[t]{2}}
\put(0.4,21){\makebox(0,0){$t_1$}}
\put(0.4,57){\makebox(0,0){$t_2$}}
\put(10.5,24){\makebox(0,0){$t_1'$}}
\put(17,46){\makebox(0,0){$t_2'$}}
\put(4.4,15){\makebox(0,0){$\varphi$}}
\put(23,39){\makebox(0,0){$A$}}
\put(0.4,8){\makebox(0,0){$O$}}
\end{picture}
\end{center}
\caption{Преобразование Лоренца.}
\label{F:l2}
\end{figure}

Как связаны координаты $x^0$, $x^1$ события $A$ с точки зрения наблюдателя 1
и координаты $x^{0\prime}$, $x^{1\prime}$ того же события с точки зрения наблюдателя 2,
мировая линия которого образует угол $\varphi$ с мировой линией наблюдателя 1?
Из рис.~\ref{F:l2} мы видим
\begin{equation*}
x^{0\prime} = \frac{t_1' + t_2'}{2}\,,\qquad
x^{1\prime} = \frac{t_2' - t_1'}{2}\,,
\end{equation*}
где
\begin{equation}
t_1' = t_1\,e^\varphi\,,\qquad
t_2' = t_2\,e^{-\varphi}\,.
\label{L:L0}
\end{equation}
Подставляя
\begin{equation*}
t_1 = x^0 - x^1\,,\qquad
t_2 = x^0 + x^1\,,
\end{equation*}
мы окончательно получаем
\begin{equation}
x^{0\prime} = x^0 \cosh\varphi - x^1 \sinh\varphi\,,\qquad
x^{1\prime} = - x^0 \sinh\varphi + x^1 \cosh\varphi.
\label{L:L}
\end{equation}
Это преобразование Лоренца.
Легко проверить, что квадрат длины вектора $x$ инвариантен:
$(x^{0\prime})^2-(x^{1\prime})^2=(x^0)^2-(x^1)^2$.
Преобразование Лоренца проще выглядит в координатах светового фронта
(приложение~\ref{S:lf}).

\section{Геометрия Минковского}
\label{S:M}

Рассмотрим вектор $x$ из точки $O$ в точку $A$.
Возможны три случая (рис.~\ref{F:cone1}):
\begin{itemize}
\item $x^2>0$ --- времениподобный вектор, может быть направлен в будущее или в прошлое
(событие $A$ лежит в будущем или прошлом события $O$).
Произведя подходящее преобразование Лоренца, можно добиться того,
чтобы единственной ненулевой компонентой была $x^0$
(для этого достаточно провести мировую линию наблюдателя
через $O$ и $A$).
\item $x^2=0$ --- светоподобный вектор, направлен вдоль светового конуса
в будущее или прошлое (событие $A$ лежит на световом конусе события $O$).
Два светоподобных вектора невозможно сравнивать (который длиннее, а который короче),
за исключением случая, когда они коллинеарны.
\item $x^2<0$ --- пространственноподобный вектор
(событие $A$ лежит в удалённом события $O$).
Произведя подходящее преобразование Лоренца, можно добиться того,
чтобы единственной ненулевой компонентой была $x^1$
(т.\,е.\ чтобы события $O$ и $A$ были одновременны).
\end{itemize}

Мы знаем, что такое квадрат вектора $x^2$~(\ref{L:x2}).
А как насчёт скалярного произведения двух векторов $x\cdot y$?
Его можно определить как
\begin{equation*}
x\cdot y = \frac{(x+y)^2 - x^2 - y^2}{2}
= x^0 y^0 - x^1 y^1\,.
\end{equation*}
В четырёхмерном пространстве--времени
$x\cdot y = x^0 y^0 - x^1 y^1 - x^2 y^2 - x^3 y^3$.
Поэтому наряду с контравариантными компонентами $x^\mu$ вектора $x$
($\mu=0$, 1, 2, 3) вводят ковариантные компоненты $x_\mu$:
\begin{equation*}
x_0 = x^0\,,\quad
x_1 = - x^1\,,\quad
x_2 = - x^2\,,\quad
x_3 = - x^3\,.
\end{equation*}
Тогда скалярное произведение имеет простой вид
\begin{equation}
x\cdot y = x^\mu y_\mu = x_\mu y^\mu\,,
\label{M:sp}
\end{equation}
где по повторяющемуся индексу (один раз сверху и один раз снизу)
всегда подразумевается суммирование от 0 до 3.

\begin{figure}[ht]
\begin{center}
\begin{picture}(43,43)
\put(21,21){\makebox(0,0){\includegraphics{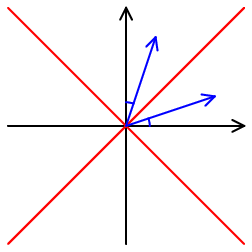}}}
\put(17,40){\makebox{$x^0$}}
\put(40,17){\makebox{$x^1$}}
\put(27,38){\makebox(0,0){$x$}}
\put(38,27){\makebox(0,0){$y$}}
\put(22.25,27){\makebox(0,0){$\varphi$}}
\put(27,21.75){\makebox(0,0){$\varphi$}}
\end{picture}
\end{center}
\caption{Ортогональные векторы.}
\label{F:ort}
\end{figure}

Векторы $x$ и $y$ называются ортогональными, если $x\cdot y=0$.
Два времениподобных вектора не могут быть ортогональны друг другу:
их скалярное произведение всегда $>0$, если оба направлены в будущее
(или в прошлое), и $<0$, если один направлен в будущее, а другой в прошлое.
Светоподобный вектор ортогонален сам себе,
а также всем коллинеарным с ним светоподобным векторам.
Он не может быть ортогонален времениподобному вектору
или светоподобному вектору, не коллинеарному с ним.
Времениподобный вектор $x$ ортогонален пространственноподобному вектору $y$,
если их направления симметричны друг другу относительно светоподобной прямой
(рис.~\ref{F:ort}):
если $y^0=x^1$, $y^1=x^0$, то $x\cdot y=0$;
ортогональность не нарушится, если $y$ умножить на скаляр.
Два пространственноподобных вектора могут быть ортогональны друг другу,
если имеется не менее двух пространственных координат:
если натянутая на них плоскость пространственноподобна,
то её геометрия евклидова, и эти два вектора могут быть ортогональны.

\begin{figure}[ht]
\begin{center}
\begin{picture}(42,28)
\put(21,13.5){\makebox(0,0){\includegraphics{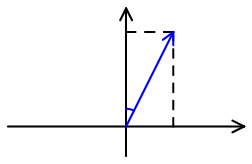}}}
\put(17,25){\makebox{$x^0$}}
\put(40,2){\makebox{$x^1$}}
\put(30,23){\makebox(0,0){$x$}}
\put(22.5,12){\makebox(0,0){$\varphi$}}
\put(20,22){\makebox(0,0)[r]{$t\cosh\varphi$}}
\put(29,5){\makebox(0,0)[t]{$t\sinh\varphi$}}
\end{picture}
\end{center}
\caption{Проекции времениподобного вектора $x$ на оси координат.}
\label{F:proj}
\end{figure}

Времениподобный вектор $x$ длины $t$ ($x^2=t^2$),
направленный под углом $\varphi$ к оси времени,
имеет компоненты~(\ref{L:csh})
$x^\mu = t(\cosh\varphi,\sinh\varphi)$
(рис.~\ref{F:proj}).
То есть его проекция на направление $e^\mu=(1,0)$
есть $t\cosh\vartheta$.
Она всегда $\ge t$; равенство достигается при $\varphi=0$.

Скалярное произведение двух времениподобных векторов $x$ и $y$
(будем считать, что оба направлены в будущее)
равно длине вектора $x$ (то есть $\sqrt{x^2}$),
умноженной на проекцию $y$ на направление $x$
(она равна $\sqrt{y^2}\cosh\varphi$):
\begin{equation}
x\cdot y = \sqrt{x^2} \sqrt{y^2} \cosh\varphi\,,
\label{M:xy}
\end{equation}
где $\varphi$ --- угол между направлениями $x$ и $y$.
Скалярное произведение всегда $\ge\sqrt{x^2}\sqrt{y^2}$;
равенство достигается, когда они коллинеарны ($\varphi=0$).

\begin{figure}[ht]
\begin{center}
\begin{picture}(104,45)
\put(21,24){\makebox(0,0){\includegraphics{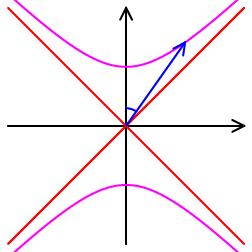}}}
\put(21,0){\makebox(0,0)[b]{a}}
\put(19,26){\makebox(0,0){$O$}}
\put(29,39.5){\makebox(0,0){$A$}}
\put(33,38){\makebox(0,0){$x$}}
\put(19.5,36){\makebox(0,0){$t$}}
\put(22.5,29){\makebox(0,0){$\varphi$}}
\put(83,24){\makebox(0,0){\includegraphics{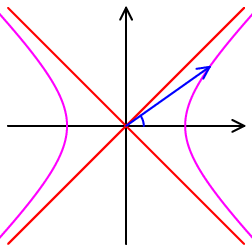}}}
\put(83,0){\makebox(0,0)[b]{b}}
\put(81,26){\makebox(0,0){$O$}}
\put(98,32){\makebox(0,0){$A$}}
\put(96,35){\makebox(0,0){$x$}}
\put(95,22){\makebox(0,0){$r$}}
\put(88,25.5){\makebox(0,0){$\varphi$}}
\end{picture}
\end{center}
\caption{``Окружности'': (a) $x^2=t^2$; (b) $x^2=-r^2$.}
\label{F:hyp}
\end{figure}

``Окружность'' --- это геометрическое место точек $A$,
удалённых от центра $O$ на расстояние $t$.
Она определяется уравнением $x^2=t^2$,
и представляет собой гиперболу (рис.~\ref{F:hyp}a).
Она имеет две ветви, в будущем и в прошлом,
представляющие собой пространственноподобные кривые;
асимптотами являются образующие светового конуса.
Верхнюю ветвь можно параметрически задать как
\begin{equation}
x^\mu = t (\cosh\varphi,\sinh\varphi)\,.
\label{M:hypt}
\end{equation}

Если же ``окружность'' имеет пространственноподобный радиус
($x^2=-r^2$, рис.~\ref{F:hyp}b),
то ветви --- времениподобные кривые,
расположенные в области удалённого.
Правая ветвь параметрически задаётся как
\begin{equation}
x^\mu = r (\sinh\varphi,\cosh\varphi)\,.
\label{M:hypr}
\end{equation}
Времениподобный вектор~(\ref{M:hypt}) ортогонален
пространственноподобному вектору~(\ref{M:hypr})
с тем же $\varphi$ (рис.~\ref{F:ort}).

\begin{figure}[h]
\begin{center}
\begin{picture}(42,42)
\put(21,21){\makebox(0,0){\includegraphics{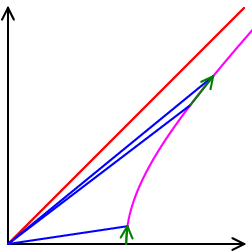}}}
\put(23,3){\makebox(0,0)[l]{$d\varphi$}}
\put(34,24.5){\makebox(0,0)[l]{$\varphi$}}
\put(37.5,29.5){\makebox(0,0)[l]{$\varphi+d\varphi$}}
\end{picture}
\end{center}
\caption{Дуги, соответствующие малых углам $d\varphi$.}
\label{F:hyp2}
\end{figure}

Точки на ``окружности''~(\ref{M:hypr}),
соответствующие углу 0 и малому углу $d\varphi$,
разделены вектором $x^\mu=(r\,d\varphi,0)$
длины $r\,d\varphi$ (Рис.~\ref{F:hyp2}).
Длина дуги между углами $\varphi$ и $\varphi+d\varphi$
равна $r\,d\varphi$ при любом $\varphi$,
потому что поворотом можно превратить угол $\varphi$ в 0,
а $\varphi+d\varphi$ в $d\varphi$.
Это легко увидеть и по-другому:
точки, соответствующие углам $\varphi$ и $\varphi+d\varphi$,
отстоят друг от друга на
\begin{equation*}
d x^\mu = (\cosh\varphi,\sinh\varphi)\,r\,d\varphi\,,
\end{equation*}
и разделены времениподобным интервалом
\begin{equation}
dx^2 = r^2 d\varphi^2\,.
\label{M:arc}
\end{equation}
То есть длина дуги с углом $d\varphi$ есть $r\,d\varphi$
(как и в евклидовой геометрии).

\begin{figure}[ht]
\begin{center}
\begin{picture}(100,51)
\put(20,27){\makebox(0,0){\includegraphics[width=40mm]{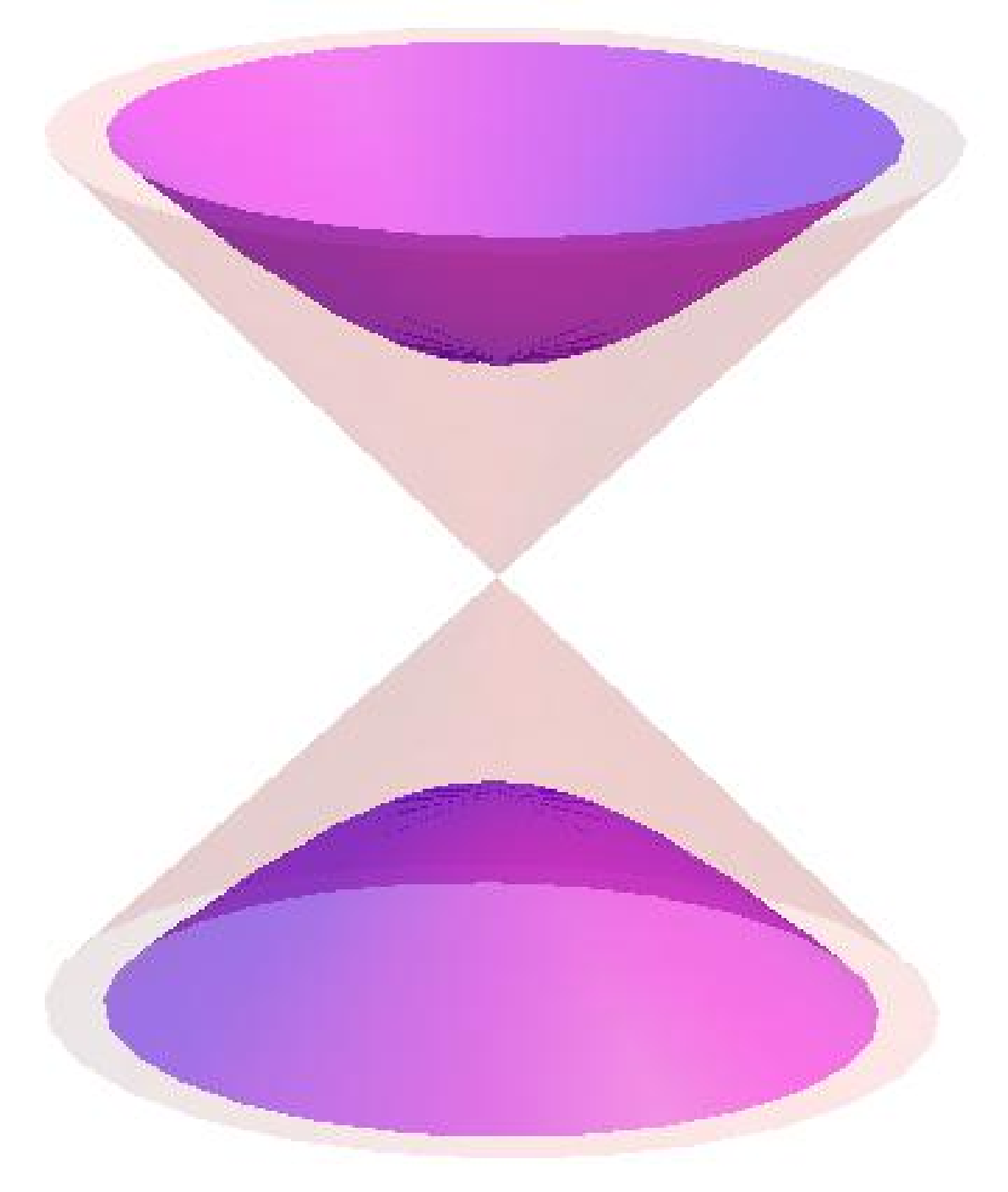}}}
\put(80,27){\makebox(0,0){\includegraphics[width=40mm]{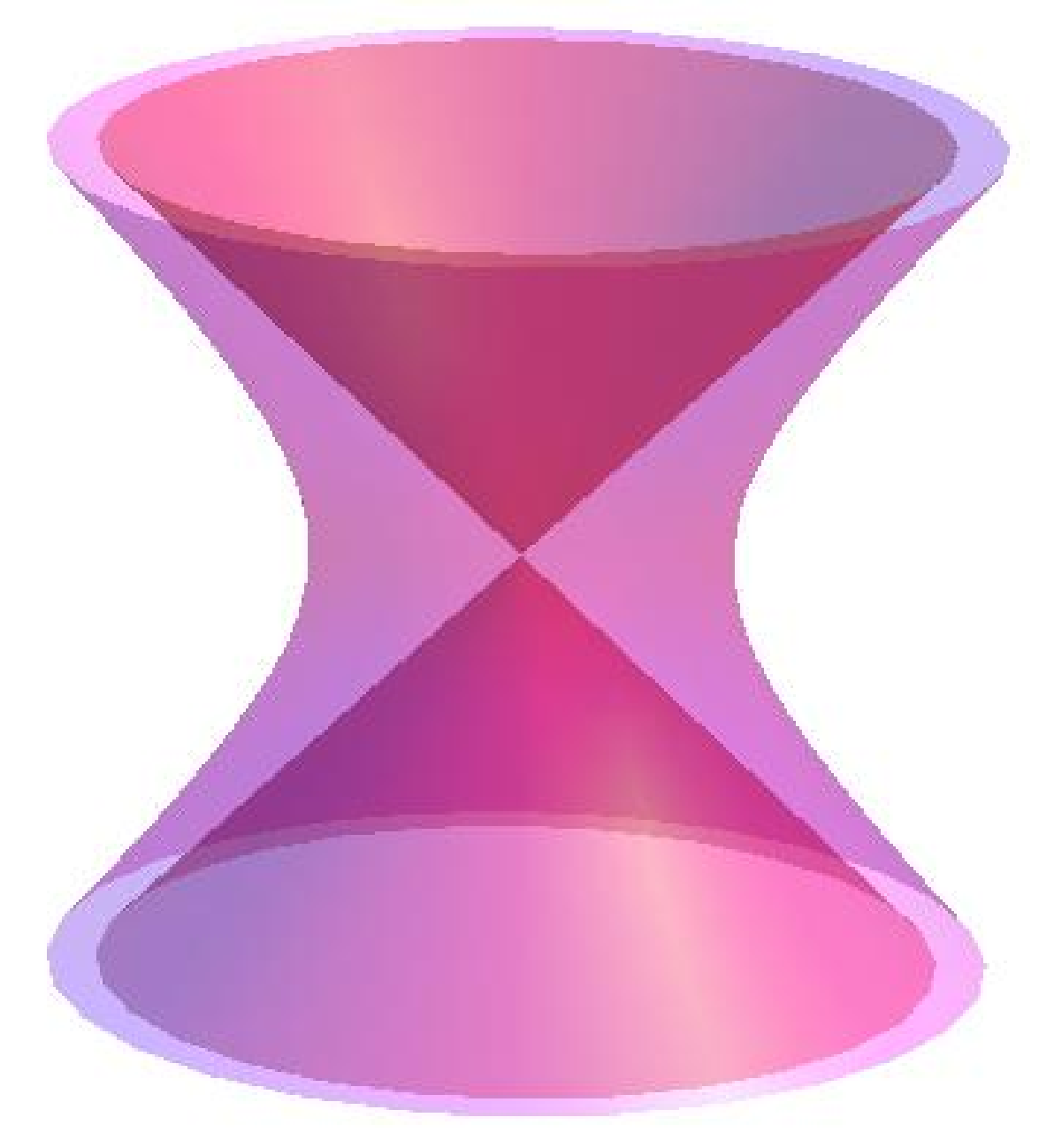}}}
\put(20,0){\makebox(0,0)[b]{a}}
\put(80,0){\makebox(0,0)[b]{b}}
\end{picture}
\end{center}
\caption{``Сферы'': (a) $x^2=t^2$; (b) $x^2=-r^2$.}
\label{F:hyp3}
\end{figure}

В трёхмерном пространстве--времени (две пространственных координаты)
``сфера'' $x^2=t^2$ представляет собой двухполостный гиперболоид (рис.~\ref{F:hyp3}a).
Он состоит из двух пространственноподобных поверхностей,
одна в будущем, а другая в прошлом.
``Сфера'' $x^2=-r^2$ --- однополостный гиперболоид (рис.~\ref{F:hyp3}b),
времениподобная поверхность, лежащая в области удалённого.
В четырёхмерном пространстве--времени их нарисовать сложнее.

\begin{figure}[t]
\begin{center}
\begin{picture}(27.75,49)
\put(15.5,24.5){\makebox(0,0){\includegraphics{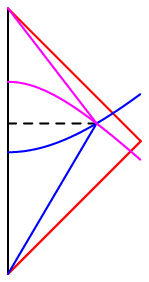}}}
\put(0,2){\makebox(0,0)[l]{$O$}}
\put(0,47){\makebox(0,0)[l]{$A$}}
\put(0,22.6){\makebox(0,0)[l]{$C$}}
\put(0,27.7){\makebox(0,0)[l]{$E$}}
\put(0,34.5){\makebox(0,0)[l]{$D$}}
\put(19.5,24.5){\makebox(0,0){$B$}}
\end{picture}
\end{center}
\caption{Неравенство треугольника.}
\label{F:tri}
\end{figure}

Рассмотрим сумму двух времениподобных векторов $x=OB$ и $y=BA$,
направленных в будущее (рис.~\ref{F:tri}).
На рисунке изображены дуги ``окружностей'':
расстояния от $O$ до $B$ и до $C$ одинаковы,
как и расстояния от $A$ до $B$ и до $D$.
Очевидно, что расстояние от $O$ до $A$ больше,
чем сумма расстояний от $O$ до $B$ и от $B$ до $A$:
\begin{equation}
\sqrt{(x+y)^2} \ge \sqrt{x^2} + \sqrt{y^2}\,.
\label{M:tri}
\end{equation}
Равенство достигается, когда $x$ и $y$ коллинеарны;
если векторы $x$ и $y$ светоподобны, то $\sqrt{x^2}+\sqrt{y^2}=0$.
Иными словами, проекция $OE$ вектора $OB$ на направление $OA$
длиннее самого вектора $OB$ (рис.~\ref{F:proj});
точно так же, проекция $EA$ вектора $BA$ на направление $OA$
длиннее самого вектора $BA$.
Это неравенство легко получить и по-другому:
достаточно возвести его в квадрат и использовать~(\ref{M:xy}).

Любую времениподобную мировую линию, соединяющую события $O$ и $A$,
можно сколь угодно точно приблизить ломаной.
Из неравенства треугольника~(\ref{M:tri}) следует,
что прямая $OA$ имеет наибольшую длину среди всех этих мировых линий.
То есть время между $O$ и $A$ по часам инерциального наблюдателя
больше, чем по часам любого другого наблюдателя,
двигавшегося с ускорением%
\footnote{Это иногда называют ``парадоксом близнецов''.
По-моему, в том, что две точки можно соединить кривыми
разной длины, нет ничего парадоксального.}.

Если у нас есть ортонормированный базис $e_0$, $e_1$,
т.\,е.\ такой, что
\begin{equation}
e_0^2 = 1\,,\qquad
e_1^2 = -1\,,\qquad
e_0 \cdot e_1 = 0\,,
\label{M:ort}
\end{equation}
то любой вектор $x$ можно по нему разложить:
\begin{equation}
x = x^0 e_0 + x^1 e_1\,.
\label{M:x}
\end{equation}
Компоненты вектора получаются проецированием на векторы базиса:
\begin{equation}
x_0 = x \cdot e_0\,,\qquad
x_1 = x \cdot e_1
\label{M:x01}
\end{equation}
(как мы уже обсуждали, $x_0=x^0$, $x_1=-x^1$).

\begin{figure}[ht]
\begin{center}
\begin{picture}(82,82)
\put(41,41){\makebox(0,0){\includegraphics{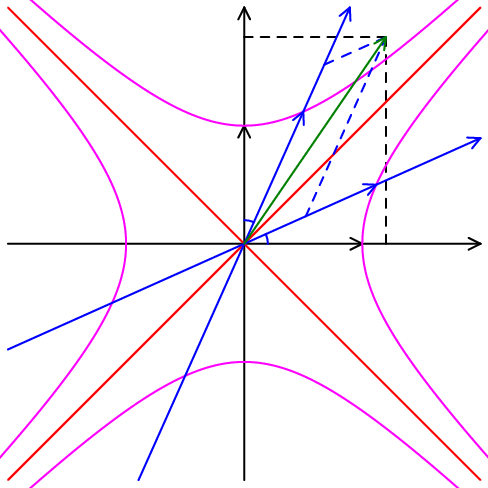}}}
\put(38.5,43){\makebox(0,0){$O$}}
\put(64.5,78){\makebox(0,0){$A$}}
\put(67,76){\makebox(0,0){$x$}}
\put(38.5,76){\makebox(0,0){$x^0$}}
\put(65,38.5){\makebox(0,0){$x^1$}}
\put(38,59.5){\makebox(0,0){$e_0$}}
\put(60,37.5){\makebox(0,0){$e_1$}}
\put(52,72){\makebox(0,0){$x^{0\prime}$}}
\put(53,44){\makebox(0,0){$x^{1\prime}$}}
\put(50,65.5){\makebox(0,0){$e_0'$}}
\put(62,53.5){\makebox(0,0){$e_1'$}}
\put(42.5,47){\makebox(0,0){$\varphi$}}
\put(47,42){\makebox(0,0){$\varphi$}}
\end{picture}
\end{center}
\caption{Преобразование Лоренца.}
\label{F:l3}
\end{figure}

В системе отсчёта инерциального наблюдателя 1,
вектор $e_0$ направлен вдоль мировой линии этого наблюдателя,
т.\,е.\ вдоль его оси времени: $e_0^\mu=(1,0)$.
Иными словами, это вектор скорости наблюдателя 1 (\S~\ref{S:v}).
Вектор $e_1$ ему ортогонален: $e_1^\mu=(0,1)$.
Пусть имеется ещё один инерциальный наблюдатель,
мировая линия которого образует угол $\varphi$
с мировой линией первого наблюдателя (рис.~\ref{F:l3}).
Орт оси времени второго наблюдателя $e_0^{\prime\mu}=(\cosh\varphi,\sinh\varphi)$
направлен из начала координат в точку на единичной окружности
под углом $\varphi$.
Орт оси $x$ ортогонален ему: $e_1^{\prime\mu}=(\sinh\varphi,\cosh\varphi)$,
и его конец лежит на окружности $x^2=-1$.
То есть орты второго наблюдателя получаются из ортов первого
поворотом на угол $\varphi$:
\begin{equation}
e_0' = e_0 \cosh\varphi + e_1 \sinh\varphi\,,\qquad
e_1' = e_0 \sinh\varphi + e_1 \cosh\varphi\,.
\label{M:rot}
\end{equation}
Компоненты вектора $x$~(\ref{M:x}) в повёрнутой системе координат
легко найти проецированием:
\begin{equation}
x_0' = x \cdot e_0' = x_0 \cosh\varphi + x_1 \sinh\varphi\,,\qquad
x_1' = x \cdot e_1' = x_0 \sinh\varphi + x_1 \cosh\varphi\,.
\label{M:L}
\end{equation}
Это ни что иное как преобразование Лоренца~(\ref{L:L}).

Некоторые примеры применения преобразования Лоренца
рассмотрены в приложениях~\ref{S:form} и~\ref{S:sky}.

\section{Скорость и ускорение}
\label{S:v}

\begin{figure}[h]
\begin{center}
\begin{picture}(42,44)
\put(21,21){\makebox(0,0){\includegraphics{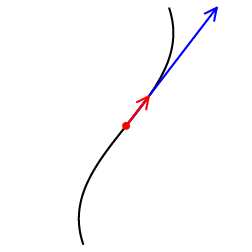}}}
\put(20,21){\makebox(0,0)[r]{$x^\mu(t)$}}
\put(22,26){\makebox(0,0)[r]{$x^\mu(t+dt)$}}
\put(25.5,26){\makebox(0,0)[l]{$dx^\mu$}}
\put(37,42){\makebox(0,0)[l]{$v^\mu$}}
\end{picture}
\end{center}
\caption{Скорость.}
\label{F:v}
\end{figure}

Мировую линию частицы можно задать параметрически как $x^\mu(t)$.
Её скорость --- это вектор
\begin{equation}
v^\mu = \frac{d x^\mu}{d t}\,,
\label{v:v}
\end{equation}
где вектор $dx^\mu=x^\mu(t+dt)-x^\mu(t)$, $dx^2=dt^2$ (рис.~\ref{F:v}).
Поэтому
\begin{equation}
v^2 = 1\,,
\label{v:v2}
\end{equation}
то есть скорость $v^\mu(t)$ ---
это единичный касательный вектор к мировой линии в точке $x^\mu(t)$,
направленный в будущее.
Мировая линия частицы, движущейся по инерции, прямая:
\begin{equation}
x^\mu = x_0^\mu + v^\mu t\,.
\label{v:x}
\end{equation}

Ускорение --- это
\begin{equation}
a^\mu = \frac{d v^\mu}{d t}\,.
\label{v:a}
\end{equation}
Поскольку $(v + d v)^2 = v^2 + 2 v\cdot d v = v^2 = 1$,
\begin{equation}
v\cdot a = 0\,.
\label{v:va}
\end{equation}
То есть ускорение --- пространственноподобный вектор,
ортогональный скорости.

\begin{figure}[h]
\begin{center}
\begin{picture}(42,42)
\put(21,21){\makebox(0,0){\includegraphics{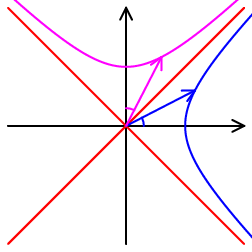}}}
\put(33,26){\makebox(0,0)[l]{$x,a$}}
\put(27,33){\makebox(0,0)[b]{$v$}}
\put(22.5,26){\makebox(0,0){$\varphi$}}
\put(26,22){\makebox(0,0){$\varphi$}}
\end{picture}
\end{center}
\caption{Равноускоренное движение.}
\label{F:a}
\end{figure}

Пусть мировой линией частицы будет правая ветвь гиперболы
\begin{equation}
x^2=-r^2
\label{v:x2}
\end{equation}
(рис.~\ref{F:a}).
Параметрически она задаётся как $x^\mu=r(\sinh\varphi,\cosh\varphi)$.
Длина дуги (т.\,е.\ время) для точки с углом $\varphi$ есть $t = r\varphi$~(\ref{M:arc})
(если выбрать начало отсчёта времени в точке $\varphi=0$).
Поскольку $(x + d x)^2 = x^2 + 2 x\cdot d x = x^2 = - r^2$,
$x\cdot v=0$, то есть скорость --- времениподобный вектор, ортогональный $x$.
При изменении $t$ от $-\infty$ до $+\infty$ конец вектора $x$
пробегает мировую линию
\begin{equation}
x^\mu(t) = r \left(\sinh\frac{t}{r},\cosh\frac{t}{r}\right)
\label{v:xt}
\end{equation}
--- правую ветвь гиперболы.
При этом конец вектора $v(t)$ пробегает верхнюю ветвь гиперболы $v^2=1$,
и в каждый момент $t$ остаётся ортогональным $x(t)$,
то есть характеризуется тем же углом $\varphi$ (рис.~\ref{F:a}).
Конечно, тот же результат можно получить, продифференцировав~(\ref{v:xt}):
\begin{equation}
v^\mu(t) = \left(\cosh\frac{t}{r},\sinh\frac{t}{r}\right)\,.
\label{v:vt}
\end{equation}
Ускорение $a^\mu(t)$ ортогонально $v^\mu(t)$, и потому направлено вдоль $x^\mu(t)$.
То же можно получить дифференцированием~(\ref{v:vt}):
\begin{equation}
a^\mu(t) = \frac{1}{r} \left(\sinh\frac{t}{r},\cosh\frac{t}{r}\right)
= \frac{x^\mu(t)}{r^2}\,.
\label{v:at}
\end{equation}
Поэтому $a^2=-1/r^2$ постоянно;
такое движение естественно назвать равноускоренным.

\begin{figure}[h]
\begin{center}
\begin{picture}(83,65)
\put(8.5,37){\makebox(0,0){\includegraphics[width=22.6182mm]{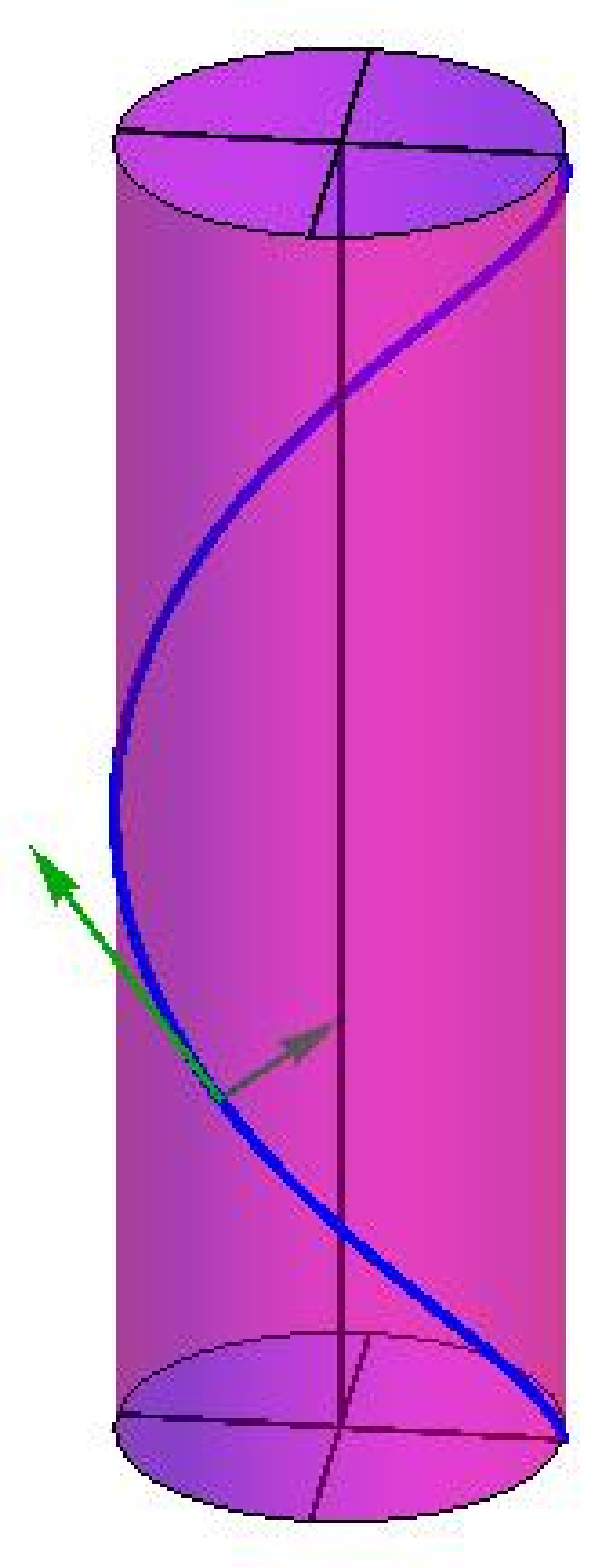}}}
\put(21.5,13){\makebox(0,0){$x^1$}}
\put(8,7){\makebox(0,0){$x^2$}}
\put(9,63.5){\makebox(0,0){$x^0$}}
\put(11.5,0){\makebox(0,0)[b]{a}}
\put(68,38){\makebox(0,0){\includegraphics[width=50mm]{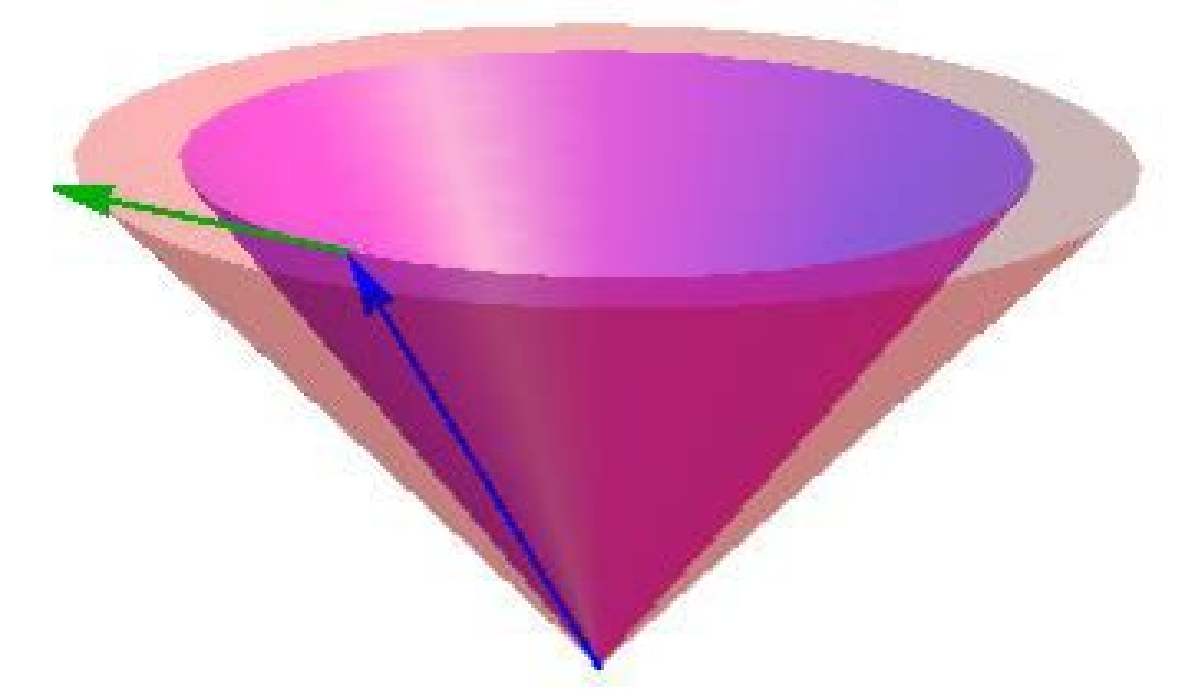}}}
\put(68,0){\makebox(0,0)[b]{b}}
\end{picture}
\end{center}
\caption{Мировая линия частицы,
равномерно движущейся по окружности.}
\label{F:spiral}
\end{figure}

Рассмотрим теперь пример с двумя пространственными координатами.
Протон в коллайдере равномерно движется по окружности радиуса $r$ с периодом $T_0=2\pi/\omega_0$
по часам покоящегося наблюдателя.
Его мировая линия --- спираль (рис.~\ref{F:spiral}a)
\begin{equation}
x^1 = r \cos(\omega_0 x^0)\,,\quad
x^2 = r \sin(\omega_0 x^0)\,.
\label{v:sp1}
\end{equation}

Если мысленно развернуть цилиндр, на который намотана спиральная мировая линия,
то станет ясно, что за период протон проходит $T_0$ вдоль оси $x^0$
и $2\pi r$ в пространственном направлении.
Поэтому период движения по часам протона равен
\begin{equation}
T = \sqrt{T_0^2 - (2\pi r)^2} = \frac{T_0}{\cosh\varphi}\,,
\label{v:spT}
\end{equation}
где
\begin{equation}
\cosh\varphi = \frac{1}{\sqrt{1 - (r \omega_0)^2}}\,,
\label{v:spphi}
\end{equation}
то есть протон движется под углом $\varphi$ к оси $x^0$.
То же можно получить дифференцированием~(\ref{v:sp1}):
\begin{equation}
dt^2 = (dx^0)^2 - (dx^1)^2 - (dx^2)^2
= (1 - r^2 \omega_0^2) (dx^0)^2\,.
\label{v:sp2}
\end{equation}
Мировую линию можно задать через собственное время $t$:
\begin{equation}
x^\mu = (t\cosh\varphi,r\cos(\omega t),r\sin(\omega t))\,,
\label{v:sp3}
\end{equation}
где $\omega = 2\pi/T = \omega_0 \cosh\varphi$
(энергия протона $E = m \cosh\varphi$, см.\ \S~\ref{S:p}).
Вектор скорости описывает конус, расположенный внутри светового конуса (Рис.~\ref{F:spiral}b):
\begin{equation}
v^\mu = (\cosh\varphi,-r\omega\sin(\omega t),r\omega\cos(\omega t))
\label{v:spv}
\end{equation}
(что получается дифференцированием~(\ref{v:sp3})).
Его конец пробегает окружность радиуса $\sinh\vartheta = \omega r$,
то есть длины $2\pi\omega r$,
за период собственного времени $T=2\pi/\omega$.
Поэтому длина вектора ускорения (пространственноподобная) равна $r\omega^2$, и
\begin{equation}
a^\mu = - r \omega^2 (0,\cos(\omega t),\sin(\omega t))
\label{v:spa}
\end{equation}
(что получается дифференцированием~(\ref{v:spv})).

\section{Импульс и волновой вектор}
\label{S:p}

Каждая частица характеризуется своей массой $m$.
Вектор
\begin{equation}
p^\mu = m v^\mu
\label{p:p}
\end{equation}
называется импульсом.
Иными словами, импульс --- это касательный вектор к мировой линии длины $m$,
направленный в будущее:
\begin{equation}
p^2 = m^2\,.
\label{p:p2}
\end{equation}
Все возможные векторы импульса частицы образуют часть ``сферы''~(\ref{p:p2}),
лежащую в будущем (рис.~\ref{F:hyp3}a),
она называется массовой поверхностью.

Безмассовые частицы имеют светоподобный импульс: $p^2=0$.
Например, фотон имеет $m=0$.
Мировые линии безмассовых частиц светоподобны.
Любой интервал времени вдоль такой мировой линии равен 0;
поэтому для них невозможно определить скорость.
Импульс, однако, хорошо определён,
и является касательным вектором к мировой линии.
Массовая поверхность частицы с $m=0$ ---
это световой полуконус будущего (рис.~\ref{F:cone}).

В системе отсчёта некоторого инерциального наблюдателя $p^\mu=(E,\vec{p})$,
где временная компонента $E$ называется энергией,
а пространственные компоненты образуют трёхмерный импульс $\vec{p}$.
Они связаны как $p^2\equiv E^2-\vec{p}\,^2=m^2$.
То есть энергия частицы, имеющей импульс $p^\mu$,
с точки зрения наблюдателя со скоростью $v^\mu$
есть $E=p\cdot v$.

Рассмотрим распад частицы с массой $m$ на две частицы с массами $m_1$ и $m_2$.
Их импульсы связаны законом сохранения
\begin{equation}
p = p_1 + p_2\,.
\label{p:cons}
\end{equation}
По неравенству треугольника~(\ref{M:tri}) $m>m_1+m_2$.
Какова энергия частицы 1 в системе покоя распадающейся частицы?
Перепишем закон сохранения~(\ref{p:cons}) в виде $p-p_1=p_2$
и возведём в квадрат:
$(p-p_1)^2=m^2+m_1^2-2p\cdot p_1=m_2^2$.
Но $p\cdot p_1=mE_1$, поэтому
\begin{equation}
E_1 = \frac{m^2+m_1^2-m_2^2}{2m}\,.
\label{p:E1}
\end{equation}
Точно так же
\begin{equation}
E_2 = \frac{m^2+m_2^2-m_1^2}{2m}\,;
\label{p:E2}
\end{equation}
легко проверить сохранение энергии $E_1+E_2=m$.
Трёхмерные импульсы продуктов распада в системе покоя $p$
противоположны: $\vec{p}_1+\vec{p}_2=0$;
по модулю они равны
\begin{equation}
\vec{p}\,^2 = E_1^2 - m_1^2 = E_2^2 - m_2^2 =
\frac{(m+m_1+m_2) (m-m_1-m_2) (m+m_1-m_2) (m-m_1+m_2)}{4 m^2}\,.
\label{p:p12}
\end{equation}
Например, в распаде $\pi^0\to\gamma\gamma$ массы фотонов $m_1=m_2=0$,
и $E_1=E_2=m/2$.

Ещё один пример --- эффект Комптона.
Фотон с энергией $E$ рассеивается на угол $\vartheta$ на покоящемся электроне.
Найдём энергию $E'$ рассеянного фотона.
Пусть импульсы начальных электрона и фотона будут $p$ и $k$,
а конечных $p'$ и $k'$:
\begin{equation}
p + k = p' + k'\,.
\label{p:cons2}
\end{equation}
Перепишем закон сохранения импульса в виде $p+k-k'=p'$
и возведём его в квадрат:
$(p+k-k')^2=m^2+2p\cdot k-2p\cdot k'-2k\cdot k'=m^2$.
В системе покоя начального электрона $p=(m,\vec{0})$,
$k=E(1,\vec{n}\,)$, $k'=E'(1,\vec{n}\,')$,
где $\vec{n}$, $\vec{n}\,'$ --- единичные трёхмерные векторы
в направлении движения начального и конечного фотонов.
Поэтому $p\cdot k=mE$, $p\cdot k'=mE'$, $k\cdot k'=EE'(1-\cos\vartheta)$,
и мы получаем
\begin{equation}
\frac{1}{E'} - \frac{1}{E} = \frac{1-\cos\vartheta}{m}\,.
\label{p:Compton}
\end{equation}

И ещё один пример.
Пион с энергией $E_\pi$ налетает на покоящийся нуклон.
При каких $E_\pi$ может происходить реакция $N \pi \to \Lambda K$
($m_\Lambda + m_K > m_N + m_\pi$)?
Закон сохранения импульса
\begin{equation}
p_N + p_\pi = p_\Lambda + p_K\,.
\label{p:cons3}
\end{equation}
Квадрат массы конечного состояния
$(p_\Lambda + p_K)^2 \ge (m_\Lambda + m_K)^2$.
Перепишем его через импульсы начальных частиц:
$(p_N + p_\pi)^2 = m_N^2 + m_\pi^2 + 2 m_N E_\pi \ge (m_\Lambda + m_K)^2$, откуда
\begin{equation}
E_\pi \ge \frac{(m_\Lambda + m_K)^2 - m_N^2 - m_\pi^2}{2 m_N}\,.
\label{p:thr}
\end{equation}
На пороге рождающиеся $\Lambda$ и $K$ имеют одинаковые скорости,
т.\,е.\ $p_\Lambda$ и $p_K$ коллинеарны.

\begin{figure}[ht]
\begin{center}
\begin{picture}(110,33)
\put(15,18){\makebox(0,0){\includegraphics{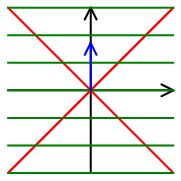}}}
\put(55,18){\makebox(0,0){\includegraphics{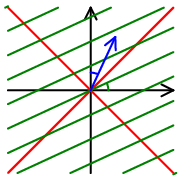}}}
\put(95,18){\makebox(0,0){\includegraphics{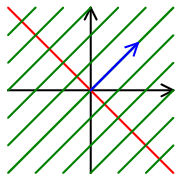}}}
\put(15,0){\makebox(0,0)[b]{a}}
\put(17,26){\makebox(0,0){$k$}}
\put(55,0){\makebox(0,0)[b]{b}}
\put(61,27){\makebox(0,0){$k$}}
\put(56,23){\makebox(0,0){$\varphi$}}
\put(60,19){\makebox(0,0){$\varphi$}}
\put(95,0){\makebox(0,0)[b]{c}}
\put(105,26){\makebox(0,0){$k$}}
\end{picture}
\end{center}
\caption{Плоские волны: волновой вектор $k$ и плоскости постоянной фазы.}
\label{F:wave}
\end{figure}

Теперь мы рассмотрим волны.
Плоская волна
\begin{equation}
e^{-ik\cdot x}
\label{p:k}
\end{equation}
характеризуется волновым вектором $k$.
Если $k^2>0$, то можно перейти в систему покоя, где $k=(k^0,\vec{0})$.
В ней волна~(\ref{p:k}) представляет собой колебания $e^{-i k^0 x^0}$,
синхронные во всём пространстве (рис.~\ref{F:wave}a).
В произвольной системе отсчёта, поверхности постоянной фазы
(например, максимумы, минимумы и нули действительной части волны~(\ref{p:k}))
представляют собой пространственноподобные плоскости,
ортогональные волновому вектору $k$ (рис.~\ref{F:wave}b).
Если $k^2=0$ (как для электромагнитных волн), то системы покоя не существует.
В этом случае плоскости постоянной фазы светоподобны,
и светоподобный вектор $k$ лежит на такой плоскости (рис.~\ref{F:wave}c).

Наблюдатель со скоростью $v$ имеет мировую линию $x=vt$.
Он воспримет волну~(\ref{p:k}) как колебания $e^{-i\omega t}$
с частотой $\omega=k\cdot v$.
Частоты волны с точки зрения разных наблюдателей различны
(эффект Допплера).

В квантовой теории нет традиционного для классической физики
различия между частицами и полями ---
квантовые поля объединяют эти понятия.
Элементарное возбуждение (квант поля),
соответствующее плоской волне~(\ref{p:k}) с волновым вектором $k$,
есть частица с импульсом
\begin{equation}
p = k\,.
\label{p:deBroglie}
\end{equation}
Это соотношение де Бройля в естественных единицах $\hbar=1$.
Так что два вектора, которые мы обсуждали в этом параграфе,
есть на самом деле одно и то же.

\appendix
\section{Часы}
\label{S:clock}

Если наблюдатель несёт с собой частицу массы $m$,
то её волновая функция зависит от времени как $e^{-imt}$.
Однако фаза волновой функции не может быть измерена,
и поэтому не может использоваться в качестве часов.

Допустим, наблюдатель взял с собой мешок
свежеприготовленных $B_s^0$ мезонов.
Они (как и $\bar{B}_s^0$ мезоны) не имеют определённой массы,
а представляют собой суперпозицию $B_{sH}^0$ и $B_{sL}^0$ мезонов:
\begin{equation}
|B_s^0{>} = \frac{|B_{sH}^0{>} + |B_{sL}^0{>}}{\sqrt{2}}\,,\qquad
|\bar{B}_s^0{>} = \frac{|B_{sH}^0{>} - |B_{sL}^0{>}}{\sqrt{2}}\,.
\label{clock:t0}
\end{equation}
$B_{sH}^0$ и $B_{sL}^0$ мезоны имеют определённые массы $m_{sH}$ и $m_{sL}$.
Поэтому по прошествии времени $t$ состояние частицы,
которая была $B_s^0$ мезоном при $t=0$, будет
\begin{equation}
\begin{split}
|B_s^0(t){>} &{}= \frac{e^{-i m_{sH} t} |B_{sH}^0{>} + e^{-i m_{sL} t} |B_{sL}^0{>}}{\sqrt{2}}\\
&{}= e^{-imt} \left[ \cos\frac{\Delta m\,t}{2}\,|B_s^0{>}
- i \sin\frac{\Delta m\,t}{2}\,|\bar{B}_s^0{>} \right]\,,
\end{split}
\label{clock:t}
\end{equation}
где $m=(m_{sH}+m_{sL})/2$, $\Delta m=m_{sH}-m_{sL}$.
То есть эта частица будет $B_s^0$ мезоном с вероятностью $\cos^2(\Delta m\,t/2)$
и $\bar{B}_s^0$ мезоном с вероятностью $\sin^2(\Delta m\,t/2)$.
Вероятности распадов, характерных для $B_s^0$ и $\bar{B}_s^0$,
будут соответственно осциллировать.
Подсчитав число этих осцилляций,
можно измерить промежуток времени между двумя событиями.

Реальные атомные часы подобным образом используют разность масс
двух стационарных состояний атома.

\section{Координаты светового фронта}
\label{S:lf}

Иногда удобно вместо компонент $x^0$, $x^1$ вектора $x$ в ортонормированном базисе
использовать
\begin{equation}
x_+ = x^0 + x^1\,,\qquad
x_- = x^0 - x^1\,.
\label{lf:def}
\end{equation}
Для события $A$ на рис.~\ref{F:l}, $x_+=t_2=t e^\varphi$, $x_-=t_1=t e^{-\varphi}$.
Квадрат вектора $x$
\begin{equation}
x^2 = x_+ x_-
\label{lf:x2}
\end{equation}
равен $t^2$.
Для времениподобного вектора $x$, направленного в будущее, $x_+>0$ и $x_->0$;
аналогично, область прошлого определяется неравенствами $x_+<0$, $x_-<0$,
а в области удалённого эти компоненты имеют противоположные знаки (рис.~\ref{F:cone1}).

Координаты $x'_+$, $x'_-$ события $A$ относительно наблюдателя 2 (рис.~\ref{F:l2})
выражаются через его координаты $x_+$, $x_-$ относительно наблюдателя 1
формулой~(\ref{L:L0}):
\begin{equation}
x'_+ = x_+ e^{-\varphi}\,,\qquad
x'_- = x_- e^\varphi\,.
\label{lf:L}
\end{equation}
При этом $x^2$~(\ref{lf:x2}) остаётся инвариантным.
Эта форма записи преобразования Лоренца эквивалентна~(\ref{L:L}),
но выглядит проще.

``Окружность'' $x^2\equiv x_+ x_-=t^2$
представляет собой гиперболу (рис.~\ref{F:hyp}a);
её верхнюю ветвь можно параметрически задать как
\begin{equation*}
x_+ = t e^\varphi\,,\qquad
x_- = t e^{-\varphi}\,.
\end{equation*}
Аналогично, правую ветвь ``окружности'' $x^2\equiv x_+ x_-=-r^2$ (рис.~\ref{F:hyp}b)
параметрически можно задать как
\begin{equation*}
x_+ = r e^\varphi\,,\qquad
x_- = - r e^{-\varphi}\,.
\end{equation*}

\begin{figure}[h]
\begin{center}
\begin{picture}(46,26)
\put(23,13){\makebox(0,0){\includegraphics{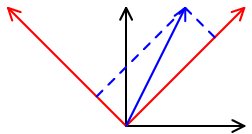}}}
\put(0,23){\makebox(0,0){$e_+$}}
\put(46,23){\makebox(0,0){$e_-$}}
\put(20,23){\makebox(0,0){$e_0$}}
\put(43,0){\makebox(0,0){$e_1$}}
\put(36,23){\makebox(0,0){$x$}}
\put(45,18){\makebox(0,0){$\frac{1}{2} x_+ e_-$}}
\put(12,6){\makebox(0,0){$\frac{1}{2} x_- e_+$}}
\end{picture}
\end{center}
\caption{Координаты светового фронта.}
\label{F:lf}
\end{figure}

Скалярное произведение~(\ref{M:sp}) в координатах светового фронта имеет вид
\begin{equation}
x\cdot y = \frac{1}{2} \left( x_+ y_- + x_- y_+ \right)\,.
\label{lf:sp}
\end{equation}
Вместо ортонормированного базиса $e_0$, $e_1$
введём базис из двух светоподобных векторов
\begin{equation}
\begin{split}
&e_+ = e_0 - e_1\,,\qquad
e_- = e_0 + e_1\,;\\
&e_0 = \frac{1}{2} \left( e_+ + e_- \right)\,,\qquad
e_1 = \frac{1}{2} \left( e_- - e_+ \right)\,.
\end{split}
\label{lf:epm}
\end{equation}
Они удовлетворяют свойствам
\begin{equation}
e_+^2 = e_-^2 = 0\,,\qquad
e_+ \cdot e_- = 2\,.
\label{lf:ort}
\end{equation}
Любой вектор $x$ можно разложить по этому базису (рис.~\ref{F:lf}):
\begin{equation}
x = \frac{1}{2} \left( x_- e_+ + x_+ e_- \right)\,;
\label{lf:x}
\end{equation}
компоненты выделяются проецированием:
\begin{equation}
x_+ = x \cdot e_+\,,\qquad
x_- = x \cdot e_-\,.
\label{lf:proj}
\end{equation}

\begin{figure}[h]
\begin{center}
\begin{picture}(52,32)
\put(26,16){\makebox(0,0){\includegraphics{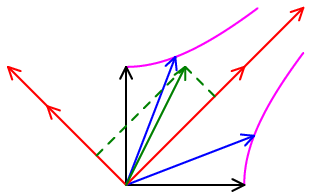}}}
\put(-1,20){\makebox(0,0){$e_+$}}
\put(5.7,13.3){\makebox(0,0){$e'_+$}}
\put(44,20){\makebox(0,0){$e_-$}}
\put(54,30){\makebox(0,0){$e'_-$}}
\put(21,23){\makebox(0,0){$e_0$}}
\put(29.3,24.7){\makebox(0,0){$e'_0$}}
\put(43,1){\makebox(0,0){$e_1$}}
\put(44.7,9.3){\makebox(0,0){$e'_1$}}
\put(33,22){\makebox(0,0){$x$}}
\end{picture}
\end{center}
\caption{Преобразование Лоренца в координатах светового фронта.}
\label{F:lf2}
\end{figure}

Если мировая линия второго инерциального наблюдателя
образует угол $\varphi$ с мировой линией первого, то
\begin{equation}
e_-' = e_- e^\varphi\,,\qquad
e_+' = e_+ e^{-\varphi}\,,
\label{lf:rot}
\end{equation}
(рис.~\ref{F:lf2}),
и компоненты вектора $x$ во второй системе отсчёта
даются формулой~(\ref{lf:L}).

\section{Видимая форма быстро движущихся тел}
\label{S:form}

Как измерить ширину коридора?
Наиболее естественно --- поперёк.
Получится некоторая характеристика этого коридора.
Но можно измерить её наискосок.
Полученная ширина характеризует не только коридор,
но и то, насколько косо её измерили.

\begin{figure}[ht]
\begin{center}
\begin{picture}(30,32)
\put(15,16){\makebox(0,0){\includegraphics{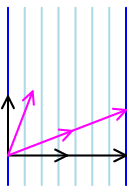}}}
\put(3,6){\makebox(0,0){$O$}}
\put(27,6){\makebox(0,0){$A$}}
\put(27,13.6923){\makebox(0,0){$B$}}
\put(3,16){\makebox(0,0){$v$}}
\put(9.3,19){\makebox(0,0){$v'$}}
\put(15,3){\makebox(0,0){$e_1$}}
\put(15,12){\makebox(0,0){$e_1'$}}
\end{picture}
\end{center}
\caption{Длина линейки с точки зрения движущегося наблюдателя.}
\label{F:Lorentz}
\end{figure}

Линейка --- это двумерная ленточка мировых линий.
Как измерить длину линейки?
Наиболее естественно --- поперёк этой ленточки,
т.\,е.\ по нормали к скорости линейки $v$ (рис.~\ref{F:Lorentz}).
То есть измерить пространственноподобный интервал
между событием $O$ на одном конце линейки
и событием $A$ на другом её конце,
одновременном с событием $O$ в системе покоя линейки.
Получится величина $l$, 
которая характеризует только эту линейку и ничего больше.
Но можно измерить её наискосок ---
между событием $O$ и событием $B$,
одновременным с $O$ для некоторого наблюдателя со скоростью $v'$.
Получится какая-то величина $l'$, которая зависит как от линейки,
так и от конкретного наблюдателя.
Проекция вектора $OB$ на ось $e_1$ имеет длину $l'\cosh\varphi = l$, так что
\begin{equation}
l' = \frac{l}{\cosh\varphi}\,.
\label{form:Lorentz}
\end{equation}
В отличие от евклидова коридора, который, если измерять его косо, становится шире,
в геометрии Минковского линейка становится короче при косом измерении.
Это называют сокращением Лоренца;
это не есть реальное физическое сокращение,
а просто следствие определения длины линейки относительно движущегося наблюдателя.

\begin{figure}[ht]
\begin{center}
\begin{picture}(74,32)
\put(16,16){\makebox(0,0){\includegraphics{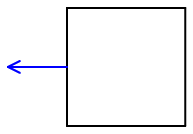}}}
\put(58,16){\makebox(0,0){\includegraphics{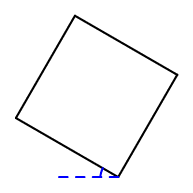}}}
\put(1,18){\makebox(0,0){$u$}}
\put(9,4){\makebox(0,0){$A$}}
\put(33,4){\makebox(0,0){$B$}}
\put(9,28){\makebox(0,0){$C$}}
\put(33,28){\makebox(0,0){$D$}}
\put(56.5,3.5){\makebox(0,0){$\alpha$}}
\put(41.6077,11.6077){\makebox(0,0){$A$}}
\put(62.3923,-0.3923){\makebox(0,0){$B$}}
\put(53.6077,32.3923){\makebox(0,0){$C$}}
\put(74.3923,20.3923){\makebox(0,0){$D$}}
\end{picture}
\end{center}
\caption{Квадратик, движущийся влево со скоростью $u$, выглядит так же,
как покоящийся, но повёрнутый на угол $\alpha$.
Наблюдатель находится снизу,
расстояние до него велико по сравнению с размером квадратика.}
\label{F:Terrell}
\end{figure}

Если мимо Вас летит какой-нибудь предмет с релятивистской скоростью,
Вы вовсе не увидите его сплющенным.
Фотоны, попадающие в глаз (или объектив фотоаппарата) в какой-то момент,
были испущены разными участками этого предмета в разные моменты времени.
Будем считать, что расстояние от наблюдателя до предмета будет много больше
размеров этого предмета.
Пусть предмет --- это квадратик со стороной $a$,
движущийся с (ньютоновской) скоростью $u$~(\ref{L:u}) (рис.~\ref{F:Terrell}).
Его сторона $AB$ будет казаться короче из-за сокращения Лоренца~(\ref{form:Lorentz}):
её длина будет $a\sqrt{1-u^2} = a/\cosh\varphi$.
Свет от стороны $CD$ был излучён на $\Delta x^0 = a$ раньше,
в это время квадратик ещё был на $au = a\tanh\varphi$ правее.
Поэтому эта сторона будет сдвинута на $au$
(и будет иметь ту же длину).
В результате движущийся квадратик будет выглядеть так же, как неподвижный,
но повёрнутый на угол $\alpha$:
\begin{equation}
\sin\alpha = u\,,\qquad
\cos\alpha = \sqrt{1-u^2}\,.
\label{form:terrell}
\end{equation}
Любой предмет можно составить из маленьких квадратиков.
Поэтому любой предмет будет выглядеть повёрнутым.
В частности, сфера будет выглядеть сферой, а не эллипсоидом.
Если размер предмета порядка расстояния до наблюдателя,
то лучи света от разных участков предмета в глаз наблюдателя
нельзя считать параллельными, и предмет будет казаться
не только повёрнутым, но и искажённым.

\section{Вид звёздного неба из быстро летящего звездолёта}
\label{S:sky}

С точки зрения наблюдателя со скоростью $v$
звезда характеризуется единичным 3-мерным вектором $\vec{n}$,
направленным от наблюдателя к звезде.
Иными словами, каждой звезде соответствуют точка $\vec{n}$
на сфере единичного радиуса --- небесной сфере.
Свет, приходящий от этой звезды, имеет светоподобный волновой вектор
\begin{equation*}
k = \omega (v - n)\,,\quad\text{где}\quad
\omega = k \cdot v\,.
\end{equation*}
Если мы рассмотрим две звезды с $n_1 = (0,\vec{n}_1)$ и $n_2 = (0,\vec{n}_2)$,
то угол $\alpha$ между направлениями на эти звёзды даётся формулой
\begin{equation}
\cos\alpha = \vec{n}_1 \cdot \vec{n}_2
= 1 - \frac{k_1 \cdot k_2}{(k_1 \cdot v) (k_2 \cdot v)}\,.
\label{sky:alpha}
\end{equation}
С точки зрения другого наблюдателя, имеющего скорость $v'$,
угол между этими звёздами равен
\begin{equation}
\cos\alpha' = 1 - \frac{k_1 \cdot k_2}{(k_1 \cdot v') (k_2 \cdot v')}\,.
\label{sky:alpha1}
\end{equation}

Пусть первый наблюдатель находится на Земле.
Направим орт $e_0$ по его скорости $v$,
а орт $e_1$ в направлении на Полярную звезду.
Тогда свет от Полярной звезды имеет
\begin{equation*}
k_1 = \omega (e_0 - e_1)\,;
\end{equation*}
свет от какой-нибудь другой звезды,
расположенной под углом $\alpha$ от Полярной в плоскости $(\vec{e}_1,\vec{e}_2)$,
имеет
\begin{equation*}
k_2 = \omega_2 (e_0 - e_1 \cos\alpha - e_2 \sin\alpha)\,.
\end{equation*}
Земной наблюдатель обнаружит, согласно формуле~(\ref{sky:alpha}),
что угловое расстояние между этими звёздами равно $\alpha$.
Второй наблюдатель пролетает мимо Земли в звездолёте
в направлении на Полярную звезду:
\begin{equation*}
v' = e_0 \cosh\varphi + e_1 \sinh\varphi\,.
\end{equation*}
Он обнаружит, согласно формуле~(\ref{sky:alpha1}),
\begin{equation*}
\cos\alpha' = \frac{\sinh\varphi + \cosh\varphi\,\cos\alpha}{\cosh\varphi + \sinh\varphi\,\cos\alpha}\,.
\end{equation*}

\begin{figure}[ht]
\begin{center}
\begin{picture}(52,42)
\put(26,21){\makebox(0,0){\includegraphics{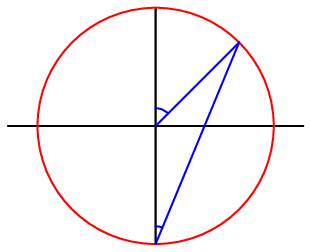}}}
\put(28,26){\makebox(0,0){$\alpha$}}
\put(27,6.5){\makebox(0,0){$\frac{\alpha}{2}$}}
\put(35,18){\makebox(0,0){$\zeta$}}
\end{picture}
\end{center}
\caption{Стереографическая проекция.}
\label{F:stereo}
\end{figure}

\begin{figure}[ht]
\begin{center}
\begin{picture}(112,90)
\put(56,69){\makebox(0,0){\includegraphics{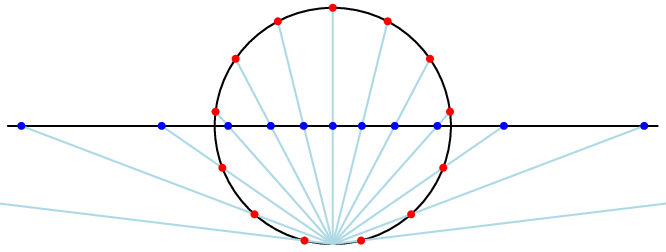}}}
\put(56,21){\makebox(0,0){\includegraphics{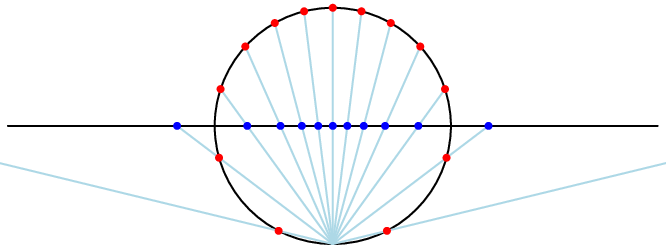}}}
\end{picture}
\end{center}
\caption{Вид звёздного неба для неподвижного наблюдателя и из быстро летящего звездолёта.}
\label{F:stars}
\end{figure}

Эту формулу можно привести к более простому виду:
\begin{equation*}
\tan\frac{\alpha'}{2} = e^{-\varphi} \tan\frac{\alpha}{2}\,.
\end{equation*}
Ей можно придать простую геометрическую интерпретацию (рис.~\ref{F:stereo}).
Будем изображать звёзды точками на плоскости,
полученными проецированием с центром на южном полюсе небесной сферы.
Звезда с углом $\alpha$ на небесной сфере
будет представлена точкой с координатой $\zeta=\tan(\alpha/2)$  на плоскости.
Полярная звезда ($\alpha=0$) соответствует $\zeta=0$;
середина Южного креста ($\alpha=\pi$) --- $\zeta=\infty$.
С точки зрения наблюдателя на звездолёте,
вся картина созвездий на плоскости равномерно сожмётся (рис.~\ref{F:stars}):
\begin{equation}
\zeta' = e^{-\varphi} \zeta.
\label{sky:zeta}
\end{equation}
На небесной сфере звёзды около Полярной звезды будут расположены гуще,
а около Южного креста реже.
Кроме того, конечно, цвет звёзд около Полярной сдвинется в синюю сторону,
а около Южного креста в красную из-за эффекта Допплера.

Если со звездолёта запустили разведывательный катер прямо вперёд,
так что его мировая линия образует минковский угол $\varphi'$ с мировой линией звездолёта,
то наблюдатель на этом катере увидит
\begin{equation*}
\zeta'' = e^{-\varphi'} \zeta' = e^{-(\varphi+\varphi')} \zeta\,,
\end{equation*}
как и следовало ожидать.

Формула~(\ref{sky:zeta}) описывает переход от покоящегося наблюдателя к движущемуся.
Если наблюдатель остался в покое,
но просто повернул голову в направлении от $\vec{e}_1$ к $\vec{e}_2$
на угол $\beta$, то $\alpha'=\alpha-\beta$;
в терминах переменной $\zeta$ это дробно-линейное преобразование
\begin{equation}
\zeta' = \frac{\displaystyle \zeta \cos\frac{\beta}{2} - \sin\frac{\beta}{2}}%
{\displaystyle \zeta \sin\frac{\beta}{2} + \cos\frac{\beta}{2}}\,.
\label{sky:frac}
\end{equation}
Полярная звезда ($\zeta=0$) теперь находится в точке $\zeta'=-\tan(\beta/2)$,
а центр Южного креста ($\zeta=\infty$) --- в точке $\zeta'=\cot(\beta/2)$.
Прямо впереди наблюдателя ($\zeta'=0$) теперь находится бывшая точка $\zeta=\tan(\beta/2)$,
а прямо сзади ($\zeta'=\infty$) --- бывшая точка $\zeta=-\cot(\beta/2)$.

До сих пор мы фактически обсуждали двумерное пространство ($\vec{e}_1$, $\vec{e}_2$),
в котором небесная сфера представляет собой окружность единичного радиуса.
Если ввести третье направление (орт $\vec{e}_3$, ортогональный $\vec{e}_1$ и $\vec{e}_2$),
то вместо прямой на рис.~\ref{F:stereo} будет плоскость.
Тогда параметр $\zeta$ надо считать комплексным.
Точки комплексной плоскости $\zeta$ (с добавлением $\infty$)
находятся во взаимно-однозначном соответствии с точками небесной сферы,
называемой сферой Римана.
Поворот наблюдателя вокруг оси $\vec{e}_1$ на угол $\gamma$
описывается преобразованием $\zeta' = e^{-i\gamma} \zeta$;
вокруг оси $\vec{e}_3$ --- формулой~(\ref{sky:frac});
а переход к движущемуся вдоль $\vec{e}_1$ наблюдателю --- формулой~(\ref{sky:zeta}).
В самом общем случае переход к другому наблюдателю даёт $\zeta'$,
являющийся дробно-линейной функцией $\zeta$.
Это преобразование конформное,
то есть в малой окрестности любой точки $\zeta_0$ оно сводится к растяжению
(или сжатию) и повороту.
То же верно для проецирования сферы Римана на комплексную плоскость
(рис.~\ref{F:stereo}).
То есть форма любого созвездия, имеющего малый угловой размер,
не изменится; оно только растянется (или сожмётся) и повернётся.
Но это растяжение--сжатие разное в разных местах небесной сферы,
так что глобальная картина звёздного неба деформируется%
\footnote{Как дробно-линейное преобразование, так и стереографическая проекция
переводят окружности в окружности, даже когда их размер не мал.
Так что если один наблюдатель проведёт окружность через какие-то звёзды,
то и для другого наблюдателя эти звёзды будут лежать на окружности.}.

\end{document}